\documentclass[a4paper,11pt]{article}

\usepackage{pos_nectarcam}
\usepackage{lipsum} %package to generate placeholder text in the following

\newcommand{\labfig}[1]{\label{fig:#1}}

\title{Timing performances of NectarCAM, a Medium Sized Telescope Camera for the Cherenkov Telescope Array}

\ShortTitle{Timing performances of NectarCAM}

\author*[1]{H. Rueda}
\author[1]{F. Bradascio}
\author[2]{J.A. Barrio}
\author[3]{J. Biteau}
\author[1]{F. Brun}
\author[4]{C. Champion}
\author[1]{J-F. Glicenstein}
\author[4]{D. Hoffmann}
\author[5]{P. Jean}
\author[6]{J.P. Lenain}
\author[1]{F. Louis}
\author[2]{A. Pérez}
\author[7]{M. Punch}
\author[1]{P. Sizun}
\author[8]{K-H. Sulanke}
\author[2]{L.A. Tejedor}
\author[1]{B. Vallage}

\affiliation[1]{IRFU, CEA, Universit\'e Paris-Saclay, F-91191 Gif-sur-Yvette, France}

\affiliation[2]{IPARCOS-UCM, Instituto de Física de Partículas y del Cosmos, and EMFTEL Department, Universidad Complutense de Madrid, E-28040 Madrid, Spain} % Uddated by JAB on 26/10/22

\affiliation[3]{Université Paris-Saclay, CNRS/IN2P3, IJCLab, 91405 Orsay, France}

\affiliation[4]{Aix-Marseille Université, CNRS/IN2P3, CPPM, 163 Avenue de Luminy, 13288 Marseille cedex 09, France}

\affiliation[5]{Institut de Recherche en Astrophysique et Plan\'etologie, CNRS-INSU, Universit\'e Paul Sabatier, 9 avenue Colonel Roche, BP 44346, 31028 Toulouse Cedex 4, France}

\affiliation[6]{LPNHE, Sorbonne Université, CNRS/IN2P3}

\affiliation[7]{Universit\'e Paris Cit\'e, CNRS/IN2P3, AstroParticule et Cosmologie (APC), Paris~F-75013, France}

\affiliation[8]{DESY, D-15738 Zeuthen, Germany}

\forColl{CTA NectarCAM} % W/O "Collaboration/Project"
\emailAdd{hector.ruedaricarte@cea.fr}
\emailAdd{federica.bradascio@cea.fr}
\abstract{

NectarCAM is a Cherenkov camera that will be installed on Medium-Sized Telescopes of the northern array of the Cherenkov Telescope Array Observatory (CTAO). It is composed of 265 modules, each of which includes 7 photo-multiplier tubes, a Front-End Board and a camera trigger system for data collection. The first NectarCAM unit is currently being integrated at CEA Paris-Saclay in France. Once installed at the CTAO's northern site, the NectarCAM's timing abilities will be crucial for reducing noise in images, improving image cleaning, and distinguishing between gamma-ray photons and cosmic-ray background. Additionally, it will enable coincidence identification with neighboring telescopes for stereoscopic observations. The timing system of NectarCAM has been tested in a dark room with various light sources. The results of the tests, including timing precision and accuracy of the trigger arrival relative to a laser source, and the timing of individual and multiple pixel signals, will be presented.
\ConferenceLogo{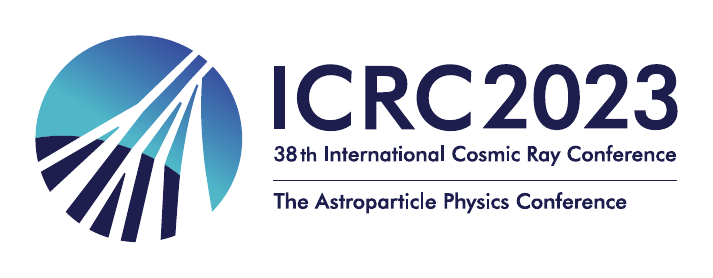}

\FullConference{The 38th International Cosmic Ray Conference (ICRC2023)\\ 26 July -- 3 August, 2023\\ Nagoya, Japan}
}

\begin{document}

\maketitle

\section{Introduction}\label{sec1}

The Cherenkov Telescope Array Observatory (CTAO)~\cite{CTAconcept, CTAdesign} is a next-generation ground-based observatory designed to detect very high-energy gamma rays from celestial sources. 
The CTAO will consist of an array of 68 imaging atmospheric Cherenkov telescopes (IACTs) located at two sites in the northern and southern hemispheres. These ground telescopes detect the faint flashes of Cherenkov radiation produced when very high energy (VHE) gamma-rays\footnote{from a few tens of GeV to above 100 TeV~\cite{CTAconcept}} from space interact with the Earth's atmosphere.
The CTAO is designed to improve the capabilities of existing gamma-ray observatories by providing a more extensive collection area and broader energy coverage~\cite{cta_mc_design}.  The array will have three size classes of telescopes: large-sized telescopes (LSTs) for low-energy gamma rays, medium-sized telescopes (MSTs) for intermediate energies, and small-sized telescopes (SSTs) for the highest energies, allowing a wide range of gamma-ray energies, from 20 GeV to 300 TeV.
NectarCAM~\cite{nectarcam} is a type of camera developed for the CTAO. It is specifically designed to be used with the medium-sized telescopes (MSTs) of the CTA, which are optimized for detecting gamma rays in the medium energy range.

The NectarCAM camera utilizes a modular design consisting of a focal plane module (FPM) and a front-end board (FEB). The FPM \cite{2021NIMPA100765413T} comprises seven Hamamatsu photomultiplier tubes (PMTs) with Winston cone light concentrators \cite{lightconcentrators}. These PMTs detect the Cherenkov light and convert it into an electric signal, which is then preamplified and split into low and high gain channels and a trigger channel in the FEB. The NECTAr chip~\cite{nectar0} plays a crucial role in sampling and digitizing the signal at a rate of 1 GHz, utilizing a 12-bit analog-to-digital converter. This chip acts as a circular buffer, holding the data until a camera trigger event occurs.

The triggering mechanism of the NectarCAM camera involves detecting a significant amount of light within a compact region of the focal plane, typically spanning around $0.2 \; \text{deg}^2$ or three pixels, within a short time frame~\cite{trigger}. Neighboring modules are triggered first (level 0 trigger or L0), followed by the camera itself (level 1 or L1, and level 2 or L2) through the combination of digitized signals at the digital trigger (DT) system. The L0 trigger is processed in each FEB using application-specific integrated circuits (ASICs), and the resulting digital signal is sent to a Field-Programmable Gate Array (FPGA) on the respective module's digital trigger backplane (DTB). The L1 trigger signal is then formed by processing the L0 signals from the module and its six neighboring modules, creating a 37-pixel region. This process ensures trigger homogeneity throughout the camera.

In evaluating the NectarCAM qualification model, particular focus was placed on timing accuracy and systematic uncertainties~\cite{BRADASCIO2023168398}.  The aim was to characterize the camera's performance by accurately timestamping triggers and estimating the arrival time of light in each pixel. These metrics are crucial in reducing noise in shower images, improving imaging cleaning, and enhancing the differentiation between Cherenkov photons and background signals.

\section{Test setup}

The first NectarCAM camera has been integrated at the CEA Paris-Saclay test facility. The facility includes a thermalized dark room with camera services and calibration light sources, connected to a control room housing the data acquisition (DAQ) \cite{daq}, storage, and control systems. The camera comprises readout electronics with 1855 pixels distributed in 265 modules and is connected to a dedicated camera server via 10 Gb/s optical links. A separate camera slow control server handles all camera devices from the control room. 

Three light sources, at a distance of 12 m from the center of NectarCAM, were used for camera evaluation: a flat field calibration light source (FFCLS) with 13 light-emitting diodes (LED); a continuous night sky background (NSB); and a laser source. The FFCLS emits pulsed light at 390 nm wavelength, reproducing the maximum of the received UV Cherenkov spectrum.
The 519 nm LED NSB source reproduces the median photoelectron rate of the typical night sky spectrum. The laser source provides uniform illumination at 373 nm wavelength and is used for precise timing measurements. 

The time of maximum (TOM) represents the reconstructed arrival time of a signal from a photomultiplier tube (PMT). The PMT signal is sampled every nanosecond within a 60-nanosecond window. By analyzing the waveform after subtracting the pedestal, the TOM is determined by finding the position of the pulse's maximum. Precise measurements of the camera's timing performance are conducted using a laser connected to a Timing and Clock-Stamping (TiCkS) board which timestamps the laser's output pulse, enabling accurate evaluation of the camera’s timing capabilities by comparison with the camera's trigger timestamp produced by another TiCkS board in the Unified Clock and Time-Stamping (UCTS) module described later.

\section{Single pixel timing precision}

In this section, we evaluate the systematic timing uncertainty of each pixel in the NectarCAM.
In order to meet the requirements of the CTAO, the NectarCAM camera needs to achieve a single pixel timing precision better than 1 ns for a light illumination above 20 photons (equivalent to 5 photoelectrons). 

We conduct experiments by illuminating the camera with a laser source, generating uniform light at frequencies of 1 kHz and intensities ranging from 8.0 nW to 20 nW. The time of maximum (TOM) of each photon pulse is measured for every pixel using two methods: identifying the largest peak position and fitting a Gaussian curve to the peak. Both methods yield consistent results within 50 ps.
The timing uncertainty of each pixel is determined by calculating the TOM distribution's root mean square (RMS).

Figure \ref{fig:pixel_res} displays the weighted mean of the TOM RMS values for all pixels as a function of the illumination charge computed from both methods. 
It shows that for light intensities above approximately 10 photons, both methods achieve a time resolution of less than 1 ns. This meets the CTAO requirement, stating that the RMS uncertainty on the mean relative reconstructed arrival time in each camera pixel should not exceed 1 ns for photon amplitudes ranging from 20 to 2000 photons per pixel. This is represented by the violet area.
However, at a light illumination of 800 photons, the pixel time precision reaches its limit (gray dashed line), corresponding to the quantified RMS noise of 1/$\sqrt{12}$ ns.

\begin{figure}[t!]
    \centering
    \includegraphics[width=0.8\textwidth]{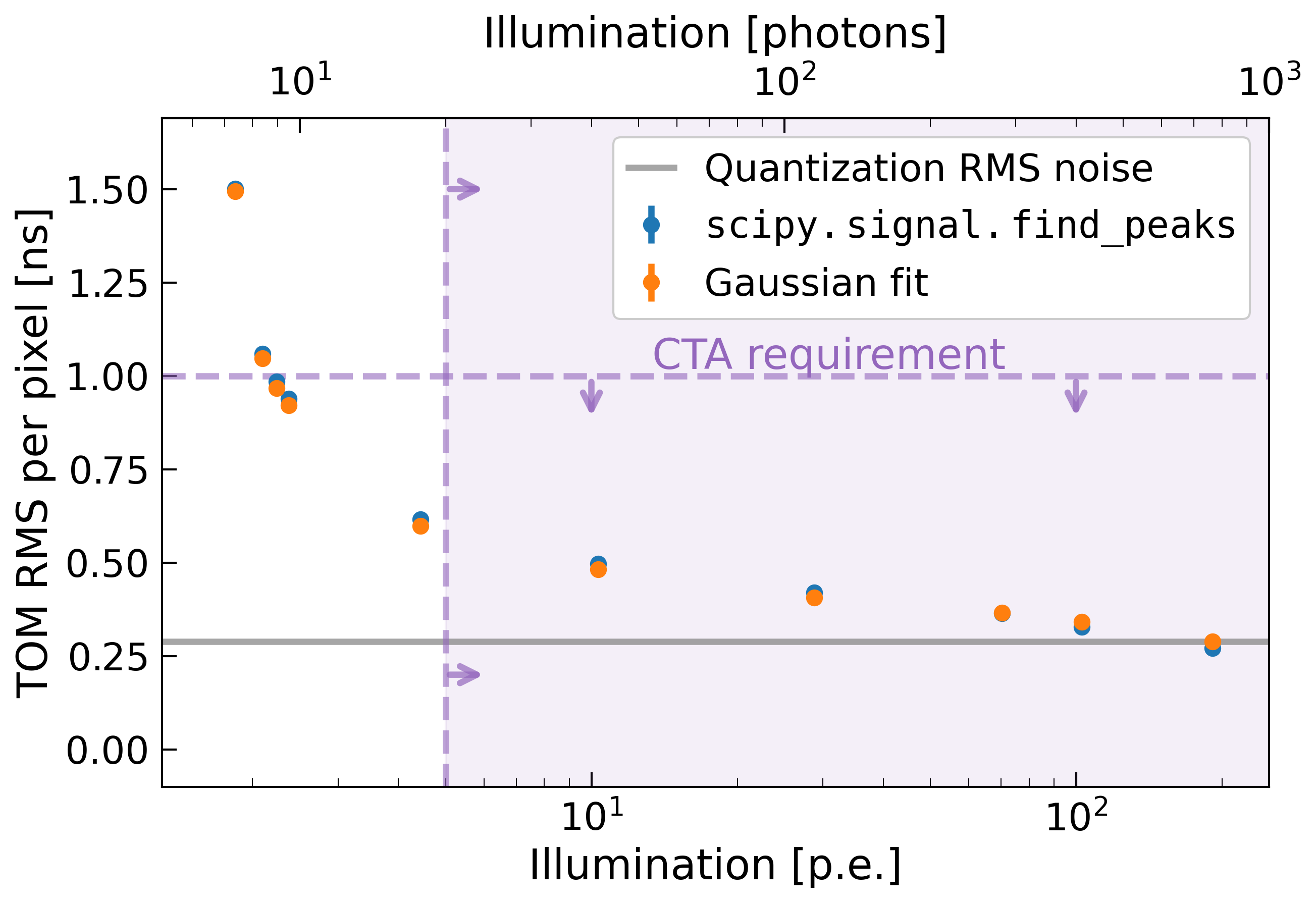}
    \caption{Timing precision per pixel (in ns) as a function of the charge of the illumination signal (in photoelectrons and photons on the bottom and top of the {horizontal} axis, respectively). The timing resolution is given by the mean of the TOM RMS distribution over all the 1855 pixels. Both methods are shown (in blue and orange). The gray {solid} line shows the {quantization} {(RMS)} noise given by $\frac{1}{\sqrt{12}}$~ns. {The dashed violet lines and arrows show} the 1~ns requirement limit to be valid between 20[5] and 2000[500]~photons~[p.e.] (violet area).}
    \labfig{pixel_res}
\end{figure}

\section{PMT transit time correction}

Prior to estimating the camera's overall timing precision, a relative calibration of the TOM is necessary to synchronize the 1855 pixels in the camera. Two systematic effects cause a timing offset between pixels. Timing offsets are influenced by variations in the arrival times of the Level 1 Accept (L1A) trigger signal sent back from the TIB due to FPGA jitter in the Digital Trigger Boards (DTBs)~\cite{ICRC2019}. Calibrating the dispersion in L1A delay involves adjusting delays in the digital backplanes and can be performed with an ordinary data-taking run.

Another effect is the PMT transit time, which can only be addressed after analysis. 
The PMT transit time, which refers to the transfer time of the electron avalanche in the PMT~\cite{leo} and depends on the high voltage applied to the dynodes, introduces delays in the camera. In order to maintain a nominal gain of 40000 across the camera's detection plane, different voltages are selected for each pixel, resulting in varying delays that typically degrade timing performances, if not corrected for. Particularly for gamma-ray energies below 100 GeV~\cite{pmt_tt_simulations}, the drop in telescope performance in terms of effective area becomes noticeable. 

Figure \ref{fig:pmt_tt} left panel illustrates the impact of transit time spread between pixels. It shows that the mean time of maximum (TOM) distributions for all pixels are not aligned but exhibit a time shift.
To correct this effect, all pixels were set to voltages ranging from 700 V to 1100 V and illuminated with various intensities from 13 LEDs. 
The dependence of each pixel's TOM as a function of their nominal voltage was evaluated by performing a least-squares fit.
Then, the PMT transit time effect was corrected by shifting the TOM values to align with the fit value at 1000 V for each pixel. 
The right panel of Figure \ref{fig:pmt_tt}  depicts the synchronized mean TOM distributions after the correction, where the PMT transit time RMS is significantly reduced.

\begin{figure*}[t!]
    \centering
    \includegraphics[width=\textwidth]{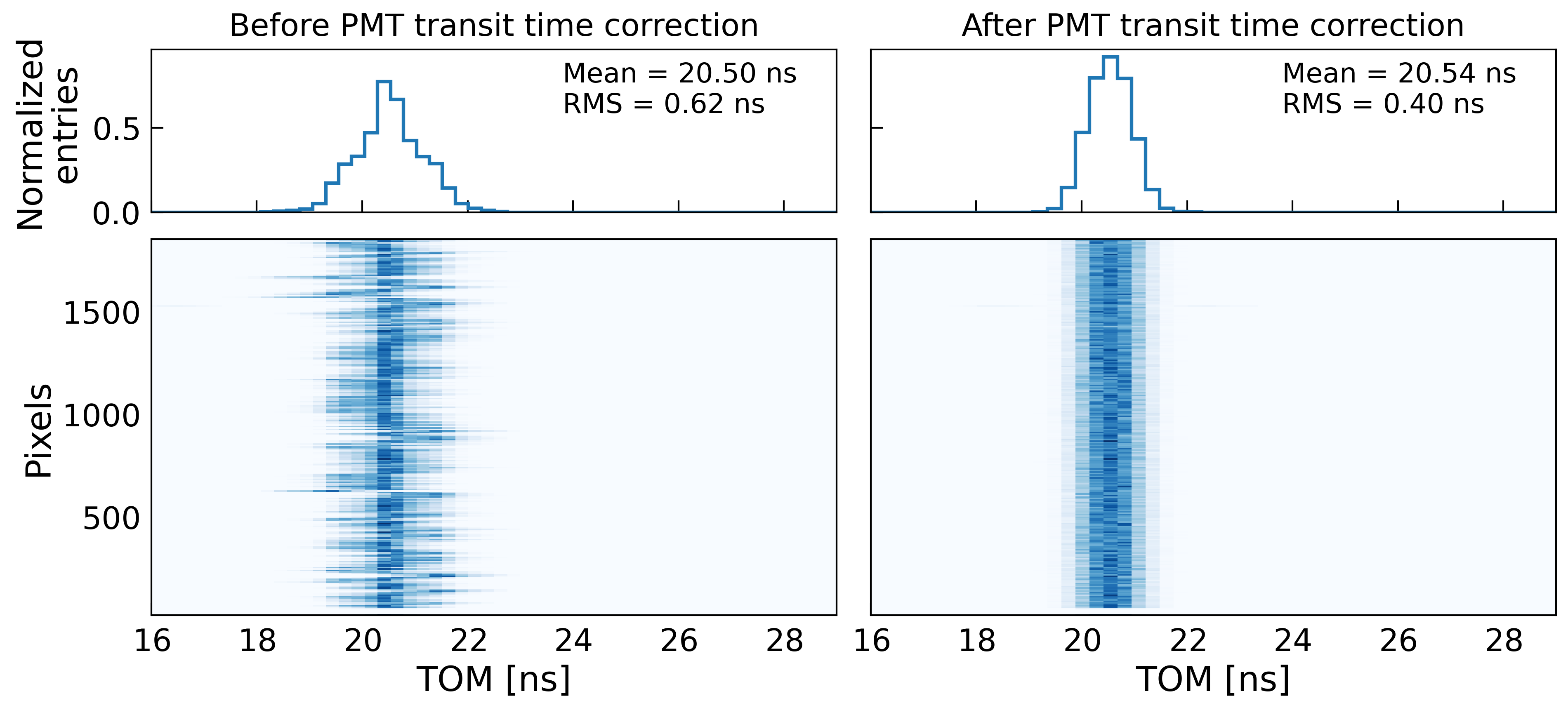}

    \caption{{Mean TOM distribution for all pixels before (left plot) and after (right plot) the PMT transit time correction for a light illumination of $\sim 70$~p.e. and for a  uniform high voltage of 1000~V for all the pixels.}}
    \labfig{pmt_tt}
\end{figure*}

\section{Global pixel timing precision}

The CTAO requirement for the RMS of the time of maximum difference distribution for any two simultaneously illuminated pixels is 2 ns.
To evaluate the timing resolution of the camera after the PMT transit time correction, we illuminate the camera with a $\sim20$ p.e. uniform light using the laser source. 

\begin{figure}[ht]
    \centering
    \includegraphics[width=0.8\textwidth]{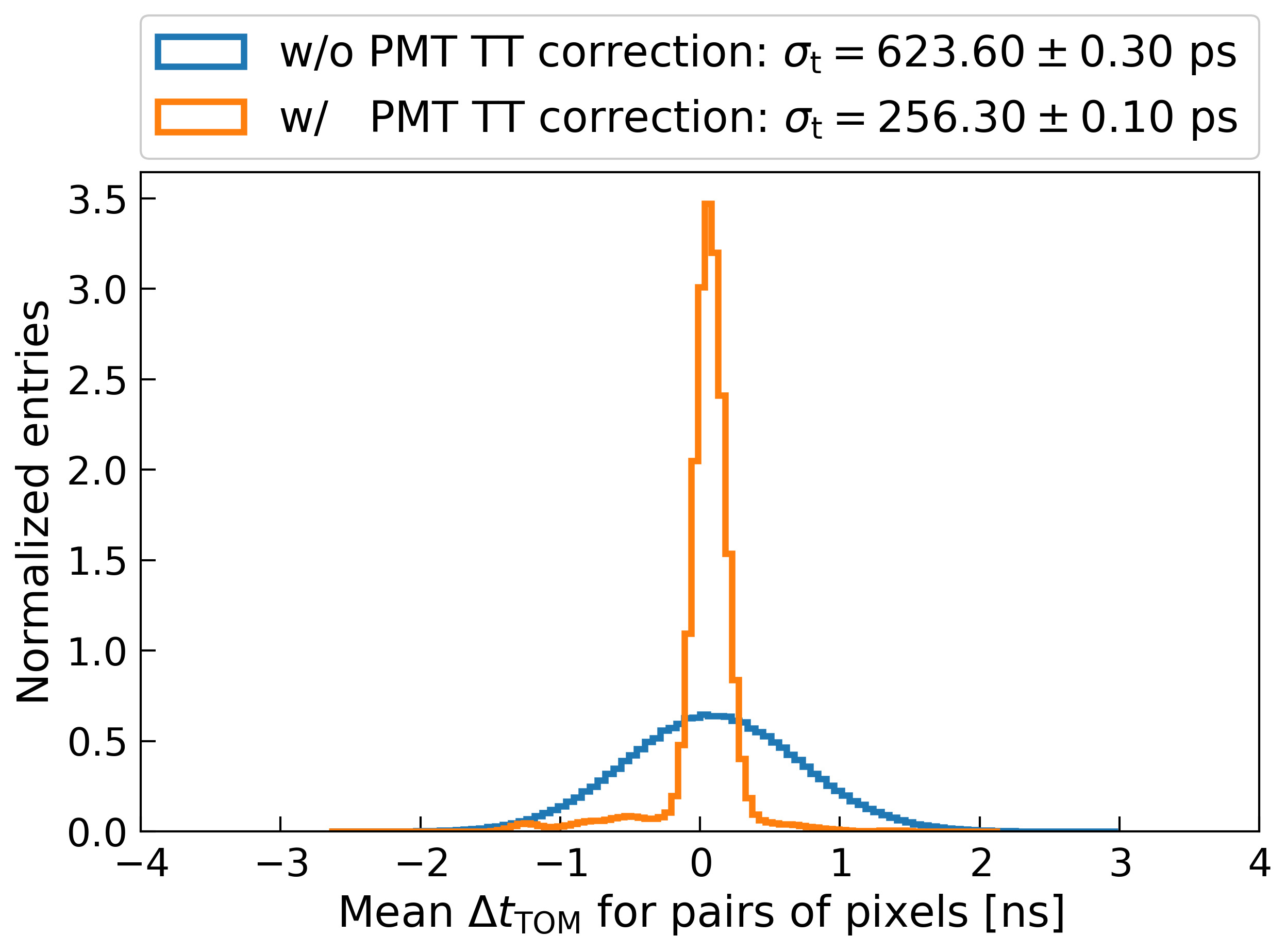}
    \caption{{Normalized distribution of the mean difference between the TOM values for each couple of pixels over all the events. The distributions with and without PMT transit time correction are shown in orange and blue, respectively. The standard deviation $\sigma_t$ of the two distributions is shown in the legend {and the error is only statistical.}}}
    \labfig{delta_t_pixels}
\end{figure}

For all pixel pairs, the TOM difference ($\Delta t_{\text{TOM}}$) is calculated with and without PMT transit time correction. Figure \ref{fig:delta_t_pixels} shows the mean distribution of $\Delta t_{\text{TOM}}$ for each pair of pixels, showing significant improvement after calibration. We see that the RMS of the $\Delta t_{\text{TOM}}$ distribution is reduced from 623.60 ps to 256.30 ps after the correction, meeting the 2~ns CTAO requirement.

\section{Camera trigger timing accuracy}

As we mentioned, the TiCkS board of the UCTS module is responsible for time-stamping the events captured by the camera. In order to ensure effective coordination and synchronization of observations across the entire CTAO array, achieving a high level of timing accuracy is crucial.
We illuminate the entire camera  using a laser source at 1 kHz with intensities ranging from 20 p.e. to 191 p.e. The start times of the laser flashes were recorded with an external TiCkS board. 
With this configuration, two different analyses are employed to quantify the RMS of the timestamp distribution. 

The first method consists in comparing the laser flashes time stamp with the light arrival UCTS in the detection plane of the camera ($\Delta t_{\mathrm{TiCkS}} = t_{\mathrm{UCTS}} - t_{\mathrm{laser}}$). 
The second analysis evaluates the distribution of the time difference between two consecutive events, expected to be $\Delta t_{\mathrm{UCTS}} = t_{\mathrm{UCTS},i} - t_{\mathrm{UCTS}, i-1} = 1 \times 10^6$~ns for a 1kHz flash. The camera's response to photons emitted by the laser, reaching the TIB and sending us its trigger pulse, is at most $\sim$245~ns, taking into account the 12~m distance between the laser and the camera's focal plane.
The RMS values obtained for each configuration are presented in Figure \ref{fig:trigger_res}. The trigger timing accuracy achieved with both methods is consistent and better than 0.5 ns, below the CTAO requirements.

\begin{figure}[t]
    \centering
    \includegraphics[width=0.8\textwidth]{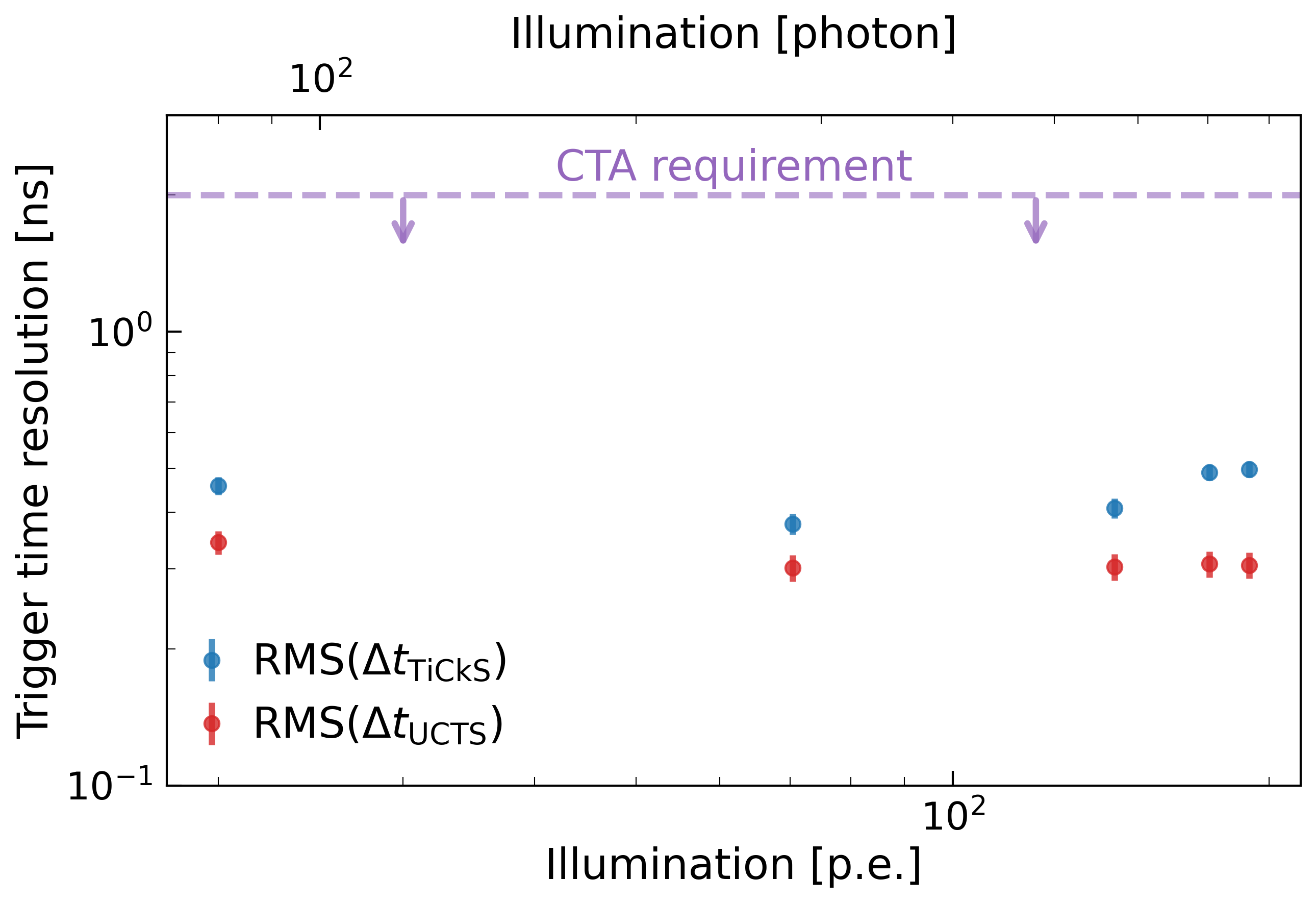}
    \caption{Camera trigger precision in ns as a function of the charge of the illumination signal (in photons and photoelectrons on the bottom and top of the {horizontal} axis, respectively) using two different methods. {The CTAO requirement of 2~ns for the camera timing resolution is also shown by the violet dashed line.} }
    \labfig{trigger_res}
\end{figure}

% As we discussed in \textsection\ref{sec1}, etc. This is a paper from a previous ICRC \cite{Zoll:2015wcu}. This is a second paper from a previous ICRC \cite{Peiffer:2017vsm}. This is a paper from the current ICRC \cite{Author:2023icrc}.
% Here is an IceCube journal paper \cite{Aartsen:2016nxy} and an external journal paper \cite{Waxman:1998yy}.

\section{Conclusion}\label{sec3}

The NectarCAM timing system has been evaluated and tested in a controlled environment, at the CEA Paris-Saclay facility. The camera has demonstrated to fullfill the CTAO timing requirements, crucial for the precise analysis of data collected. The experiments on single-pixel timing precision show that the camera achieves a time resolution of less than 1 ns for light intensities above 10 photons. The PMT transit time calibration effectively reduces the timing offsets between pixels, improving the overall timing precision of the camera. After calibration, the global pixel timing precision satisfies the CTAO requirement of 2 ns for the RMS of the TOM difference distribution between any two simultaneously illuminated pixels. Additionally, the camera trigger timing accuracy has been evaluated using a laser source, and the results demonstrate that the trigger timing is better than 0.5 ns, surpassing the CTAO requirements. 

\section*{Acknowledgments}
This work was conducted in the context of the CTA Consortium. We gratefully acknowledge financial support from the agencies and organizations listed here: \url{https://www.cta-observatory.org/consortium_acknowledgments/}.

% Bibtex references:
\bibliographystyle{ICRC}
\bibliography{references}

% Alternatively, you can include references by hand:
%\begin{thebibliography}{99}
%\bibitem{...}
%
%\end{thebibliography}

% \clearpage

%The following list of authors, affiliations and funding agencies will be updated at the day of submission.
%\input{authorlist_HESS.tex}

%The following template is a placeholder generated via https://authorlist.icecube.wisc.edu/icecube on March 25, 2023 and will be updated.
% \input{authorlist_IceCube.tex}

\end{document}